\documentclass[journal=jacsat,manuscript=article]{achemso}

\usepackage[version=3]{mhchem} 

\author{Swayang Priya Mahanta}
\affiliation{Laboratory for Nanomagnetism and Magnetic Materials (LNMM), School of Physical Sciences, National Institute of Science Education and Research (NISER), an OCC of Homi Bhabha National Institute (HBNI), Jatni 752050, Odisha, India}
\author{Antarjami Sahoo}
\affiliation{Laboratory for Nanomagnetism and Magnetic Materials (LNMM), School of Physical Sciences, National Institute of Science Education and Research (NISER), an OCC of Homi Bhabha National Institute (HBNI), Jatni 752050, Odisha, India}
\author{Sagarika Nayak}
\affiliation{Laboratory for Nanomagnetism and Magnetic Materials (LNMM), School of Physical Sciences, National Institute of Science Education and Research (NISER), an OCC of Homi Bhabha National Institute (HBNI), Jatni 752050, Odisha, India}
\author{T. P. A. Hase}
\affiliation{Department of Physics, University of Warwick, Coventry, CV4 7AL, United Kingdom}
\author{Del Atkinson}
\email{del.atkinson@durham.ac.uk}
\affiliation{Department of Physics, Durham University, South Road, Durham DH1 3LE, United Kingdom}
\author{Subhankar Bedanta}
\email{sbedanta@niser.ac.in}
\affiliation{Laboratory for Nanomagnetism and Magnetic Materials (LNMM), School of Physical Sciences, National Institute of Science Education and Research (NISER), an OCC of Homi Bhabha National Institute (HBNI), Jatni 752050, Odisha, India}
\alsoaffiliation[Second University]
{Center for Interdisciplinary Sciences (CIS), National Institute of Science Education and Research (NISER), An OCC of Homi Bhabha National Institute (HBNI), Jatni, Odisha 752050, India}

\title[An \textsf{achemso} demo]
  {Spinterface Mediated Magnetic Properties of Co$_{20}$Fe$_{60}$B$_{20}$/Alq$_{3}$ Heterostructures}

\abbreviations{IR,NMR,UV}

\begin{document}


\begin{abstract}
Organic semiconductors (OSCs) are suitable materials for spintronics applications as they form a spinterface when placed next to a ferromagnet, which in turn leads to novel functionalities. The evolution of spinterface can tune the global magnetic anisotropy, magnetization reversal, magnetization dynamics etc. Planar 
tris(8-hydroxy-
quinoline)aluminium (Alq$_{3}$) OSC has shown tremendous potential for spintronics application, thanks to its efficient spin-polarized current transport ability. Here, we establish the spinterface when the Alq$_{3}$ molecules are deposited on amorphous ferromagnet Co$_{20}$Fe$_{60}$B$_{20}$(CFB). The $\pi$-$d$ hybridization in CFB/Alq$_{3}$ enhances the coercive field and significantly modifies the shape and size of the magnetic domains. A$\sim$ 100$\%$ increase in uniaxial anisotropic energies and a reduction in magnetic damping are also evident owing to the strong interfacial hybridization. 
\end{abstract}

\section{Introduction}
Spintronics based on metallic and metal oxide multilayered thin-films is well established in terms of applications for data storage in hard disk drives and magnetic RAM and research continues to further our understanding of spintronic physics. In contrast, organic spintronics is a more recent emerging field that applies principles of spintronics to systems incorporating organic materials, offering new opportunities for spin-based devices, processing routes and applications\cite{wolf2001spintronics,naber2007organic,sanvito2011molecular,pandey2023perspective}. By utilizing the unique properties of organic materials, such as their flexibility, low-cost fabrication, and tuneable electronic and spin properties, organic spintronics aims to bridge the gap between conventional inorganic spintronics and organic electronics. The field started with the discovery of giant magnetoresistance of 40 $\%$ at 11K with Alq$_3$ as a spacer layer in a spin valve structure with LSMO and Co as ferromagnetic metal electrodes \cite{barraud2010unravelling}. The organic semiconductors (OSCs), composed of low Z elements, exhibit low spin-orbit coupling, less hyperfine interaction, and a long spin lifetime. Hence, the OSCs have proven to be good candidates for the transport of spin-polarized signals in organic spintronics devices\cite{zhang2020application,guo2019spin}. In the last decade, the OSCs have also been found to be useful in interface tailoring to modulate magnetic properties such as magnetization reversal, magnetization dynamics, magnetic anisotropy, and to modify the magnetic domain structure in ferromagnet (FM)/OSC heterostructures\cite{mallik2018effect,mallik2019enhanced,sharangi2021spinterface,bairagi2015tuning}. 
\par
When organic molecules are deposited onto a ferromagnet, a range of interactions take place from weak physisorption to strong chemisorption \cite{pandey2023perspective,delprat2018molecular}. In an isolated organic molecule, the density of states is discrete and equal for both spin-up and down electrons, in contact with a ferromagnet, the $d$-orbital of the ferromagnet hybridizes with the $p$-orbital of the OSC modifying the density of states (DOS) of OSC and the FM. In addition to molecular hybridization, charge transfer from the FM to the OSC can also take place at the interface\cite{pandey2023perspective,delprat2018molecular}. The hybrid interface between the FM and OSC has been termed as “spinterface”\cite{sanvito2010rise}. Several reports have shown the impact of such hybridization on the overall anisotropy and magnetization reversal, mainly with fullerene-based spinterface \cite{mallik2018effect,mallik2019enhanced,sharangi2021spinterface,sharangi2022effect,bairagi2015tuning}. However, the possibility of spinterface formation with other promising OSCs and their impact upon the magnetization dynamics and domain structure engineering is yet to be fully explored. Among OSCs, Alq$_{3}$ (tris(8-hydroxyquinoline)aluminum) is well-known because of its abundant potential benefits in organic spin valves\cite{sun2010giant,zhang2014spin}. Alq$_{3}$ consists of a central aluminium atom bonded to three molecules of 8-hydroxyquinoline (Figure \ref{fig:Schematic} (a))\cite{colle2004thermal}. This coordination complex gives Alq$_{3}$ its unique electronic and photophysical properties. It has been explored as a potential material for spintronic devices due to its unique properties, such as strong electron-phonon coupling, long spin coherence times, and efficient spin injection capabilities\cite{droghetti2016dynamic}.
Dediu et al. were the first to achieve room-temperature spin valve operation in an organic spin valve with Alq$_{3}$ as the spacer layer\cite{dediu2008room}. They observed approximately -9$\%$ magnetoresistance (MR) signal at low temperatures and about -0.15$\%$ MR at room temperature. Similarly, Xiong et al. have also observed a giant magnetoresistance of -40$\%$ at 11 K by taking Alq$_{3}$ as spacer layer\cite{xiong2004giant}. It is also a key organic compound that has revolutionized organic light emitting diode (OLED) technology\cite{tang1987organic,fukushima2012green}. The exceptional physical properties of Alq$_{3}$ have made it a widely utilized material for displays and other electronic devices. The work function of FMs is typically high (around 5.0 eV) which is closer to the highest occupied molecular orbital (HOMO) level of Alq$_{3}$ (around 5.8 eV)\cite{jang2012observation}. Hence, the HOMO levels of Alq$_{3}$ hybridize with the $d$-orbitals of FM. Zhan et al., have studied the spin injection through exchange coupling at the (Alq$_{3}$)/FM(Co and Fe) heterojunction\cite{zhan2010efficient}. Their findings suggest a strong coupling between Fe and Co with Alq$_{3}$ at the interface. X-ray magnetic circular dichroism (XMCD) measurement also reveals the presence of magnetic interaction between Alq$_{3}$ and the ferromagnetic metal atoms, which showed that the Alq$_{3}$ molecules hybridize and become magnetized when deposited on a ferromagnet\cite{zhan2010efficient}. These fundamental developments in understanding invite questions about the implications and impact of such interfacial hybridization on the functional magnetic properties that include anisotropy, magnetic domain structure, magnetization reversal, dynamics and damping. A detailed study on magnetization reversal, magnetic domains, magnetic anisotropy, and magnetic damping on the FM/Alq$_{3}$ system is presented in this manuscript. Specifically, we show the stabilization of the spinterface for Alq$_{3}$ deposited on the technologically important amorphous ferromagnetic Co$_{20}$Fe$_{60}$B$_{20}$ (CFB). The interactions across the interface in the CFB/Alq$_{3}$ heterostructure strongly modify the magnetic anisotropy and the magnetization dynamics of the system. A significant reduction in domain size in CFB thin films when interfaced with Alq$_{3}$ is also observed. These exotic spinterface physical phenomena can be explained by the  $\pi$-$d$ hybridization evident at the FM/OSC the interfaces\cite{bairagi2015tuning,kawahara2012large,zhan2010efficient}.
\section{Experimental details}
\begin{figure}
\includegraphics[scale=0.5]{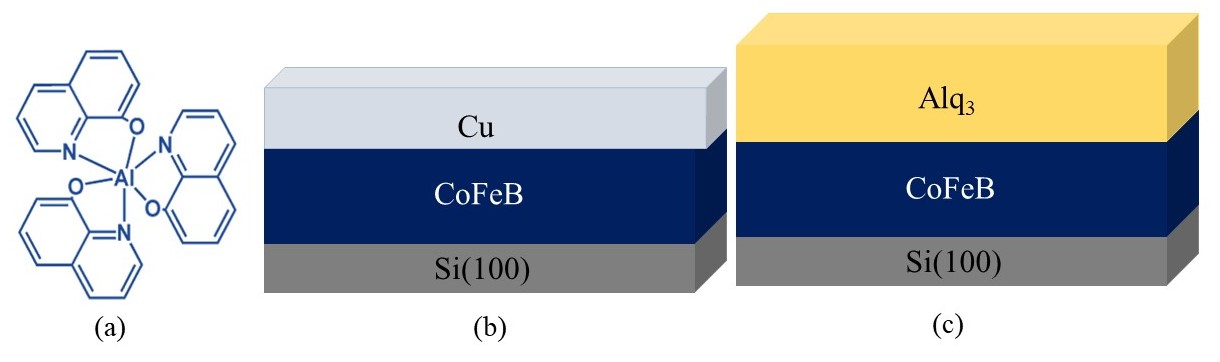}
\caption{(a) Chemical structure of Alq$_{3}$ molecule,(b) and (c) represent schematics of sample structures with and without Alq$_{3}$ molecules.}
\label{fig:Schematic}
\end{figure}

Heterostructures of CFB/Alq$_{3}$ were prepared by a combination of sputtering and thermal evaporation. The sample structure and nomenclature are listed in Table \ref{tbl:Sample}.
\begin{table}
  \caption{Details of the hetero-structures along with their nomenclature.}
  \label{tbl:Sample}
  \begin{tabular}{l|l}
    \hline
    Stacking  & Nomenclature  \\
    \hline
    Si(100)/ CFB(6 nm)/Cu(3 nm)   & SA1   \\
    Si(100)/ CFB(6 nm)/Alq$_{3}$(14 nm) & SA2  \\
    Si(100)/ CFB(10 nm)/Cu(3 nm)  & SB1  \\
    Si(100)/ CFB(10 nm)/Alq$_{3}$(14 nm) & SB2 \\
    \hline
  \end{tabular}
\end{table}
DC magnetron sputtering was employed to grow the metallic layers (CFB and Cu), while the  organic Alq$_{3}$ layer was deposited by thermal evaporation using an effusion cell in a cluster-deposition high vacuum chamber manufactured by Excel Instruments, India. The base pressure of the sputtering and evaporation chambers were 4 $\times$ 10$^{-8}$ and 6 $\times$ 10$^{-9}$ mbar, respectively. The deposition pressure for the metals was $\sim$10$^{-3}$ mbar and $\sim$10$^{-8}$ mbar for the Alq$_{3}$ evaporation. The Alq$_{3}$ layer was deposited onto the CFB layer without breaking the vacuum, to avoid surface contamination. In addition, a single-layer Alq$_{3}$ film was grown for structural characterization. During deposition, the substrate was rotated at 20 revolutions per minute (rpm) to improve the uniformity of the layer growth. The rate of deposition of CFB, Alq$_{3}$, and Cu were 0.15, 0.14, and 0.05 \AA/s, respectively. The thin Cu layer was used as capping layer to prevent oxidation of the CFB hetero-structures without  Alq$_{3}$. In-situ film thickness and deposition rates were monitored using a gold-coated quartz sensor crystal. During all the depositions, the substrate was kept at room temperature. Schematics for the hetero-structures CFB/Cu and CFB/Alq$_{3}$ are shown in Figure \ref{fig:Schematic} (b) and (c).
Structural characterization of the hetero-structures was performed using X-ray reflectivity (XRR) with a Rigaku four-cycle X-ray diffractometer and Raman spectrometry manufactured by Horiba Scientific with a 532 nm laser.  Magnetic domain imaging and magnetization reversal were performed by magneto-optic Kerr effect (MOKE) based microscopy manufactured by Evico magnetics GmbH, Germany. Magnetization dynamics were studied using ferromagnetic resonance (FMR) spectroscopy with a coplanar waveguide (CPW) based system (manufactured by NanOsc, Sweden). FMR measurements were performed with a DC magnetic field and FMR spectra were measured in the frequency range of 3-17 GHz at intervals of 0.5 GHz. Each spectrum was analysed to obtain the FMR linewidth ($\Delta H$) and resonant field ($H_{res}$) values that were used to evaluate the Gilbert damping parameter ($\alpha$). Further to quantify the effective magnetic anisotropy, in-plane angle ($\phi$) dependent FMR was measured at 7 GHz through angle from 0$^\circ$ to 360$^\circ$ with a step of 5$^\circ$. 
\section{Results and discussion}
\begin{figure}
\includegraphics[scale=0.6]{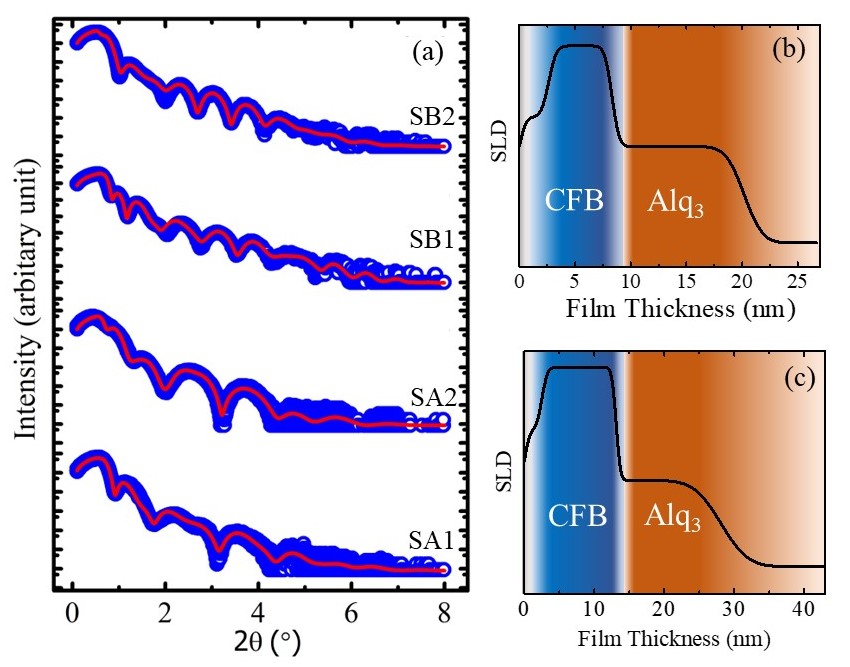}
\caption{(a)XRR data and the corresponding best fits for all the hetero-structures. The blue circles represent the experimental data and the red curve represents the best fits using GenX software. SLD plots of (b) SA2 and (c) SB2 hetero-structures.}
\label{fig:XRR}
\end{figure}

\begin{figure}
\includegraphics[scale=0.4]{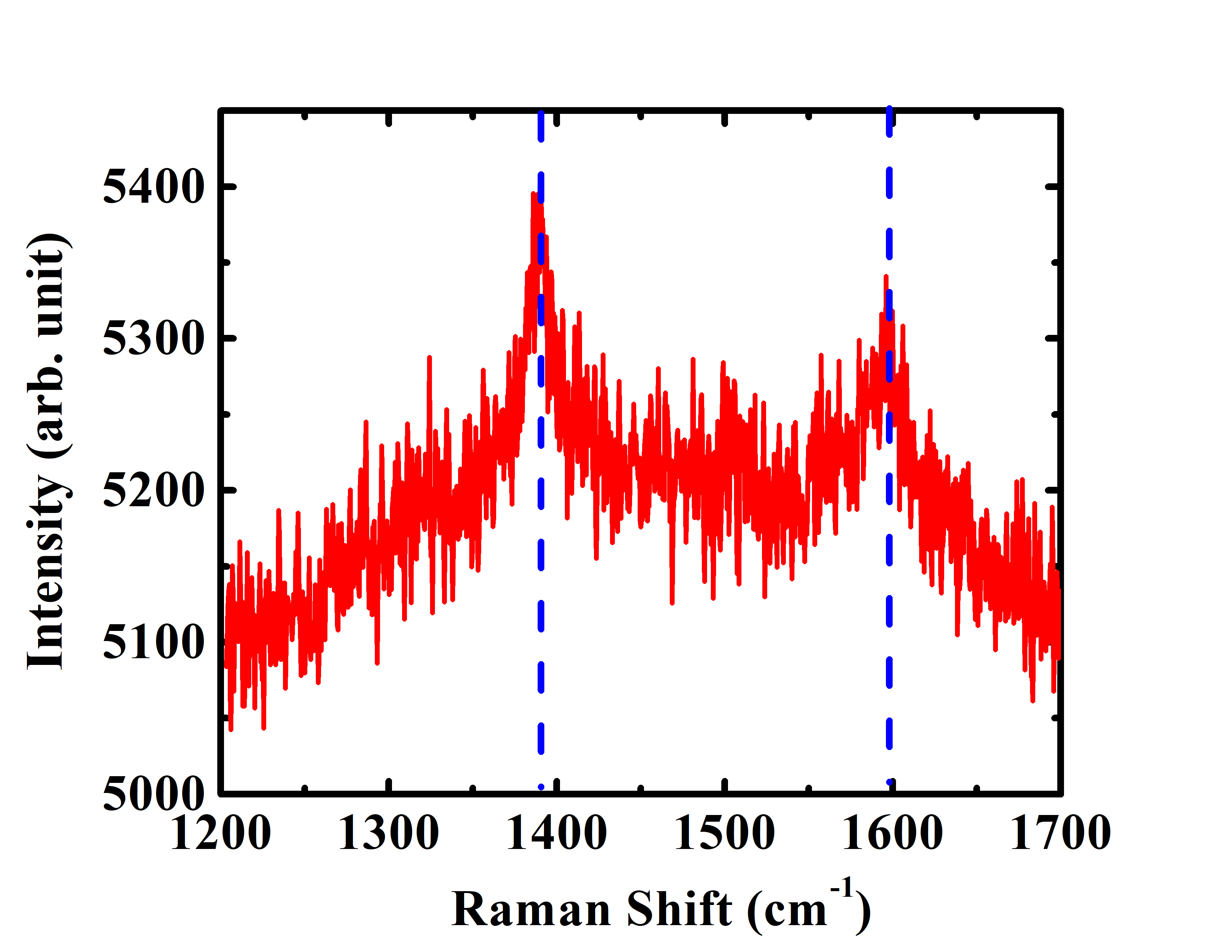}
\caption{Raman spectrum of 14 nm thick Alq$_{3}$ film deposited on (100) Si substrate}
\label{fig:Raman}
\end{figure}

Considering first the structural analysis, XRR measurements were performed for the thin films and bilayers to investigate the layer thicknesses, effective density and the extent of the interface between the layers. Specular reflectivity plots and the corresponding best fitting simulated profiles, obtained using GenX software\cite{glavic2022genx} are shown in Figure \ref{fig:XRR} (a). The thicknesses of each layer extracted from the fit are in agreement with the nominal thicknesses of the CFB, Alq$_{3}$, and Cu deposited layers mentioned in Table \ref{tbl:Sample}. The fits also indicate that the width of the interfaces for different hetero-structures is in the range of 0.2-0.5 nm. The scattering length density (SLD) plots simulated via GenX for SA2 and SB2 hetero-structures are presented in Figure \ref{fig:XRR} (b-c). The electron density for both CFB and Alq$_{3}$ is found to be uniform throughout the individual layer thickness with a sharp transition around the CFB/Alq$_{3}$ interface, inferring the good quality of the interfaces.
The Raman spectrum from a 14 nm thick Alq$_{3}$ film grown on a Si substrate is shown in Figure \ref{fig:Raman}. The intensities of the peaks are very low because of the low thickness of Alq$_{3}$ in the film. We observe two vibrational modes at $\sim$1390 cm$^{-1}$ and $\sim$1597 cm$^{-1}$, which are similar to the previously reported Raman modes of Alq$_{3}$\cite{galbiati2015molecular}. The Raman mode at $\sim$1390 cm$^{-1}$ is associated with the aromatic C=C stretching mode and $\sim$1597 cm$^{-1}$ corresponds to the C=N stretching mode in the quinoline ring. The Raman mode at $\sim$1597 cm$^{-1}$ also corresponds to the bending of C-H bonds\cite{sakurai2004study,davis2009surface}. The Raman intensity is suppressed when Alq$_{3}$ layers are deposited over the CFB layer because of the presence of free electrons in the metallic CFB layer, which cannot show polarizability change.
\par
\begin{figure}
\includegraphics[scale=0.6]{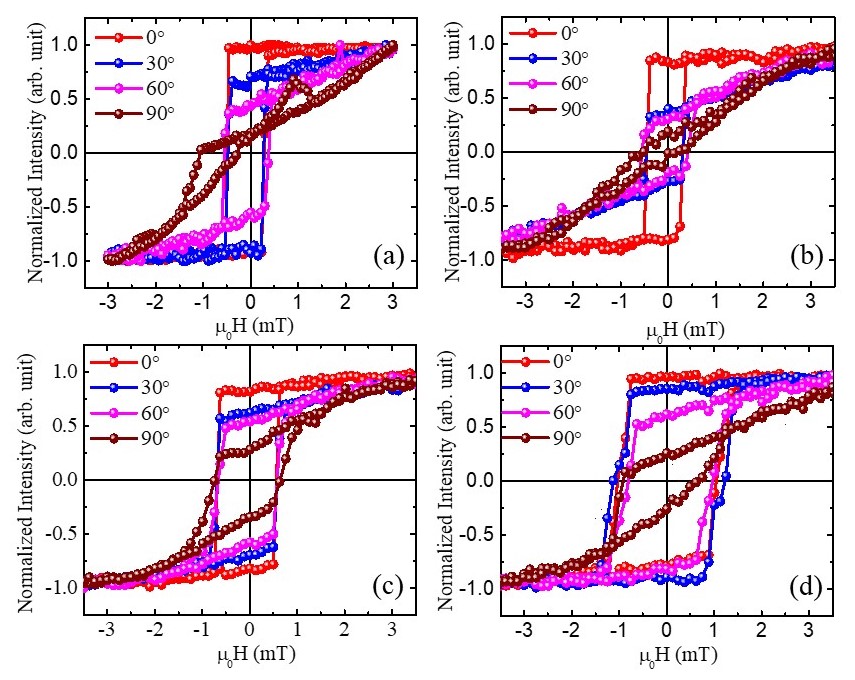}
\caption{The hysteresis loops for the hetero-structures (a) SA1, (b) SA2, (c) SB1, and (d) SB2 recorded at room temperature by varying the angle ($\phi$) between the easy axis and the applied magnetic field direction.}
\label{fig:Hys}
\end{figure}
\begin{figure}
\includegraphics[scale=0.5]{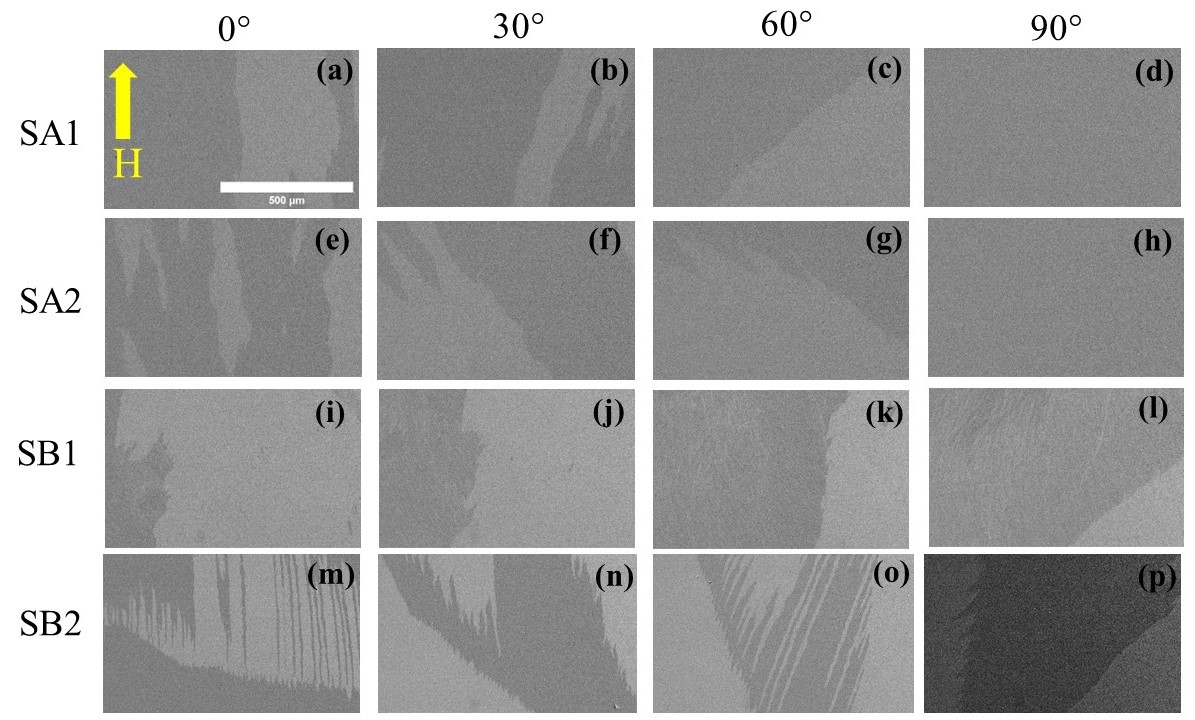}
\caption{Domain images recorded via MOKE microscopy at room temperature near coercive field for hetero-structures SAI, SA2, SB1, and SB2 are shown in (a-d), (e-h), (i-l), and (m-p), respectively. The scale bar of the images for all the hetero-structures are the same and is shown in (a). The applied field direction shown in (a) is kept same during the angle ($\phi$) dependent measurements for all the hetero-structures.}
\label{fig:Domain}
\end{figure}

Considering now the magnetic hysteresis and the anisotropy, the in-plane angle ($\phi$) dependent magnetic hysteresis loops were obtained from MOKE microscopy for all hetero-structures, see Figure \ref{fig:Hys}. In all cases a clear angular dependence is present. Square loops with a remanence ratio around 1.0 are evident for all hetero-structures when the magnetic field is applied along the angle defined as $\phi$ = 0$^\circ$    indicating an easy axis. Moving away from this axis of orientation of the magnetic field, the remanent magnetization starts decreasing and consequently, the squareness of hysteresis loops gets reduced and approaches hard axis behaviour for $\phi$ $\sim$ 90$^\circ$.   This type of hysteresis loop behaviour indicates the presence of a uniaxial magnetic anisotropy in the system.  The square-shaped loops along the easy axis indicate the magnetization reversal is dominated by domain wall motion as indicated by MOKE microscopy, whereas, along and near the hard axis, the reversal is dominated by coherent rotation\cite{hubert1998domain,hubert1998magnetic}. This uniaxial anisotropy in heterostructures may be ascribed to the oblique angle of sputtered deposition of CFB\cite{smith1960oblique}, as in the deposition chamber all the target cusps are at an angle $\sim$ 35$^\circ$ with respect to the substrate because of its in-build geometry. This oblique angle deposition can result in a preferential direction of magnetic anisotropy in the ferromagnetic film. 
However, using the MOKE technique, quantification of magnetic anisotropy is difficult. Thus, it is quantified with the help of the FMR technique which is discussed later in this article. Further, the coercive field ($H_c$) has been found to be enhanced for SB2 as compared to its reference SB1 hetero-structure by 0.48 mT, which can be attributed to the interfacial hybridization. Such behavior is not observed for SA1 hetero-structure as the crystallinity of the 6 nm CFB film can be different compared to the 10 nm thick CFB film. A similar type of coercive field modification with organic molecules has also been previously observed in other FM/OSC hetero-structures\cite{mallik2018effect,mallik2019enhanced,mallik2019tuning,sharangi2021spinterface}. The angle-dependent loop shapes of SA2 and SB2 also get modified compared with the hetero-structures without Alq$_{3}$, which indicates the modification in magnetic anisotropy because of the formation of spinterface. 
The domain images near $H_{c}$ for all the hetero-structures are shown in Figure \ref{fig:Domain}. The angle ($\phi$) mentioned in the top row is the angle between the easy axis and the applied magnetic field direction and the sample names are mentioned in the left column. In general, larger magnetic domains are observed for the reference CFB/Cu hetero-structures i.e. in SA1 and SB1. Interestingly, we observe the reduction in magnetic domain size for SA2 and SB2 relative to their corresponding reference SA1 and SB1 hetero-structures, respectively. However, the effect is more prominent in SB2 sample compared to that with the SA2. The reduction of magnetic domain size can be explained due to the formation of spinterface at the CFB and Alq$_{3}$ interface. The prominent modification in magnetic domains in SB2 is also consistent with the enhanced coercivity as seen in the magnetization reversal measurements. Branched as well as stripe domains are observed in the hetero-structures with organic Alq$_{3}$ layers. It is known that the stripe domains are characteristics of the films involving dispersed uniaxial anisotropy, \cite{hubert1998domain,hubert1998magnetic,mallik2019tuning} whereas the branched domains usually appear for films with amorphous growth\cite{hubert1998domain,hubert1998magnetic,sharangi2021spinterface}.
Interestingly, the branching becomes significant when the Alq$_{3}$ is deposited on top of CFB, indicating a significant change in the microstructure of the FM layer. The prominent modification in domain shape and size for SB2 hetero-structure upon the deposition of Alq$_{3}$ makes it a potential candidate for domain engineering via spinterface stabilization. 
\par
\begin{figure}
\includegraphics[scale=0.65]{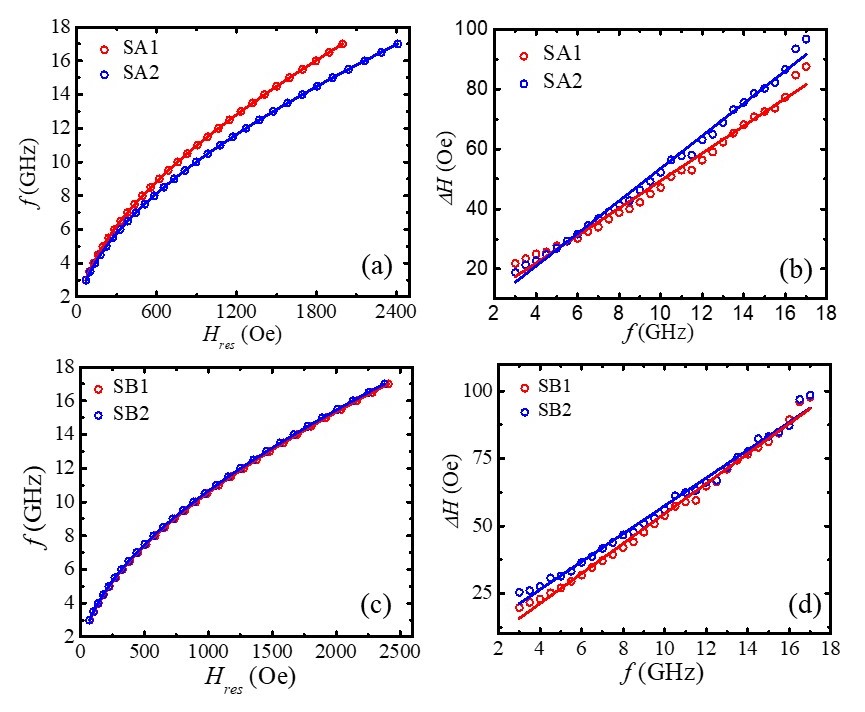}
\caption{(a) $f$ vs. $H_{res}$ plots for samples SA1 and SA2, (b) $\Delta H$ vs. $f$ plots for samples SA1 and SA2, (c) f vs. $H_{res}$ plots for samples SB1 and SB2, and (d) $\Delta H$ vs. $f$ plots for samples SB1 and SB2. The circles in the plot represent the experimental data and solid lines are the best fits.}
\label{fig:kittel}
\end{figure}
\begin{figure}
\includegraphics[scale=0.5]{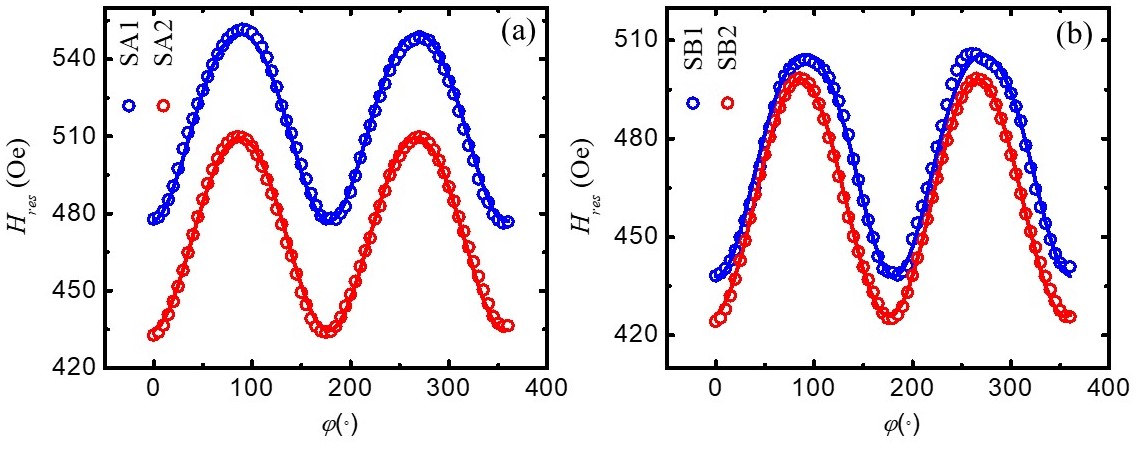}
\caption{Angle dependent resonance field ($H_{res}$) plots for hetero-structures (a) SA1 and SA2, and (b) SB1 and SB2. The measurements were performed at a constant frequency of 7 GHz at room temperature.}
\label{fig:Anisotropy}
\end{figure}

The magnetization dynamics of all the heterostructures were investigated by the FMR-based measurements. FMR spectra recorded in the 3-17 GHz range of all the heterostructures. The FMR response at each frequency was fitted with a Lorentzian function (equation \ref{eq:FMR}) \cite{woltersdorf2004spin}:
\begin{equation}
    FMR Signal = A_1 \frac{4(\Delta H)(H-H_{res})}{[(\Delta H)^2 + 4 (H - H_{res})^2]^2} - A_2 \frac{(\Delta H)^2 - 4(H-H_{res})^2}{[(\Delta H)^2 + 4 (H - H_{res})^2]^2} + Offset,
    \label{eq:FMR}
\end{equation}
where, $\Delta H$, $H_{res}$, $A_1$, and $A_2$ are the line width, resonance field, antisymmetric, and symmetric components, respectively.  From each fit, $H_{res}$ and $\Delta H$ are obtained which were used to estimate the values of $\alpha$. The $f$  vs. $H_{res}$ and $\Delta H$ vs. $f$  plots for all the hetero-structures are shown in Figure \ref{fig:kittel}.  The  $f$ vs. $H_{res}$ plot is fitted using equation \ref{eq:Kittel} (Kittel equation) to obtain the values of Gyromagnetic ratio ($\gamma$), anisotropy field ($H_K$), effective demagnetization field (4$\pi M_{eff}$) etc \cite{kittel1948theory,kalarickal2006ferromagnetic}.
\begin{equation}
    f = \frac{\gamma}{2 \pi} \sqrt{(H_K + H_{res})(H_K +H_{res}+4\pi M_{eff})},
    \label{eq:Kittel}
\end{equation}
where  $$\gamma= (\frac{g\mu_B}{\hbar})$$ in which $g$, $\mu_B$, $\hbar$ are the Lande g-factor, Bohr magneton, the reduced Planck’s constant, respectively.  Values for the inhomogeneous linewidth broadening($\Delta H_0$) and the Gilbert damping parameter ($\alpha$) were obtained from the $\Delta H$ vs. $f$ plot using equation \ref{eq:damping}\cite{heinrich1985fmr}.
\begin{equation}
    \Delta H = \Delta H_0 + \frac{4 \pi \alpha}{\gamma}f,
    \label{eq:damping}
\end{equation}
where the  $\Delta H_0$ is the inhomogeneous linewidth broadening.
The effective Gilbert damping parameter and the $\Delta H_0$ values are listed in Table \ref{tbl:MagPar}. The low $\Delta H_0$ values for all the heterostructures infer the absence of any significant artefacts and imperfections\cite{swindells2022interface}.  The $\alpha$  value reduces when the CFB layer is covered by the Alq$_{3}$ molecules, compared to that with the CFB/Cu bilayers. This reduction in magnetic damping is interesting. The enhancement of damping is known to come from layering with larger spin-orbit coupling (SOC) material and consequently, from spin-pumping. A reduction may suggest some change of the SOC at the interface with the Alq$_{3}$ or may reflect a reduction due to the absence of Cu, which can enhance the damping a small amount due to spin pumping. Achieving anti-damping behaviour in magnetic materials is often envisaged as it enhances the efficiency of the device's performance.  Further, the observance of anti-damping with organic molecules is quite rare in literature. Our investigations suggest that the Alq$_{3}$ modulates the magnetization reversal phenomena and can also act as an active component in spintronics devices for tuning the magnetic damping of ferromagnetic thin films.  
The magnetic anisotropy has been quantified by performing in-plane angle ($\phi$) dependent FMR measurement at 7 GHz. Measurements were performed by rotating the sample with respect to the external magnetic field, at 5$^\circ$ intervals. The $H_{res}$ vs. $\phi$ plots are shown in Figure \ref{fig:Anisotropy}. 
The plots are fitted with the dispersion relation for the resonance condition in the in-planegeometry by equation \ref{eq:anisotropy}, for the evaluation of anisotropic constants\cite{pan2017role}.
\begin{equation}
    f = \displaystyle\frac{\gamma}{2\pi}\sqrt{((H + \frac{2K_2}{M_S} Cos(2\phi))(H + 4\pi M_S + \frac{2K_2}{M_S} Cos^2 (\phi)))} 
    \label{eq:anisotropy}
\end{equation}
where, $K_{2}$ is the in-plane uniaxial anisotropy constant,  $\phi$ is the in-plane angle between the easy axis and the applied magnetic field direction, and $M_S$ is the saturation magnetization. The $K_2$ values obtained from the fittings are listed in table \ref{tbl:MagPar}.
\begin{table}
  \caption{$\alpha$, $\Delta H_0$ and $K_2$ values for all the hetero-structures extracted from the fitting using equation \ref{eq:damping} and \ref{eq:anisotropy}.}
  \label{tbl:MagPar}
  \begin{tabular}{l|l|l|l}
    \hline
    Hetero-structures  & $\alpha$ & $\Delta H_0$ (Oe) & $K_2$ ($\times$10$^3$ erg/cc)\\
    \hline
SA1 & 0.0074 $\pm$ 0.0002 & 4 $\pm$ 1 & 12.7 $\pm$ 0.1\\
SA2 & 0.0069 $\pm$ 0.0001 & 0 $\pm$ 1 & 23.7 $\pm$ 0.2 \\
SB1 &  0.0072 $\pm$ 0.0001 & -1 $\pm$ 1  & 20.2 $\pm$ 0.2 \\
SB2 & 0.0061 $\pm$ 0.0001  & 6 $\pm$ 1 & 23.1 $\pm$ 0.1 \\
    \hline
  \end{tabular}
\end{table}
$K_2$ increases when the CFB layer is covered by Alq$_{3}$ molecules in the SA2 and SB2 hetero-structures compared to that with their reference SA1 and SB1 hetero-structures. The enhancement of anisotropic energy in the system can be attributed to the spinterface formation of Alq$_{3}$ with CFB. It should be noted that the enhancement of $K_2$ is quite significant for the SA2 heterostructure compared to SB2. This indicates the important role of FM layer thickness in realizing different spinterface effect\cite{mallik2019tuning}. The degree of crystallization and orbital orientations in hetero-structures with 6 and 10 nm of CFB can be different. Hence, the spinterface behaviour which mainly arises due to the hybridization of $\pi$ orbitals of OSCs with various $d$-orbitals of FMs can be governed by the thickness of FM layers, the relative orientations of different $\pi$ and $d$ orbitals etc. The gigantic enhancement in magnetic anisotropy energy without much change of the coercive field for SA2 hetero-structure compared to SA1 is quite interesting and further theoretical understanding is required to unveil its deep-rooted physics. However, a similar effect has also been reported for CFB/C$_{60}$ hetero-structure.\cite{sharangi2021spinterface} The significant modification in magnetic domain sizes, coercive fields, magnetic anisotropic energy, and magnetic damping owing to the CFB/Alq$_{3}$ hybridized interface is quite important for intriguing fundamental spinterface physics and future spintronics device applications.  
\section{Conclusion}
In this work, we have investigated the magnetization reversal and magnetization dynamics of CFB/Alq$_{3}$ heterostructures. The spinterface gets stabilized when the Alq$_{3}$ molecules are stacked on the amorphous ferromagnetic CFB layer. The in-plane magnetization reversal measurements reveal the magnetic hardening of CFB owing to the interfacial hybridization. The spinterface induces a significant reduction in magnetic domain sizes of CFB which can pave the way for efficient domain engineering with organic molecules. Further, the magnetization dynamics study of CFB/Alq$_{3}$ bilayers infers a reduction in damping compared to CFB/Cu bilayers. The strong interfacial hybridization also results in a nearly 100$\%$ increase of uniaxial anisotropic energy that can be tuned by the thickness of the CFB layer.  Our investigations suggest that Alq$_{3}$ offers the possibility for tailoring magnetic domains and magnetic anisotropy. Hence, Alq$_{3}$ can also be considered as a prospective candidate for creating power-efficient spin-based electrical devices with potential benefits over conventional inorganic materials.
\begin{acknowledgement}
The authors acknowledge the financial support by the Indo-UK project funded by the Royal Society, United Kingdom. We also acknowledge the Department of Atomic Energy (DAE), the Department of Science and Technology (DST-SERB) of the Government of India. A.S. acknowledges the DST-National Postdoctoral Fellowship in Nano Science and Technology. We are thankful to the Center for Interdisciplinary Sciences, NISER for providing Raman spectroscopy measurement facility.
\end{acknowledgement}
\bibliography{achemso-demo}

\providecommand{\latin}[1]{#1}
\makeatletter
\providecommand{\doi}
  {\begingroup\let\do\@makeother\dospecials
  \catcode`\{=1 \catcode`\}=2 \doi@aux}
\providecommand{\doi@aux}[1]{\endgroup\texttt{#1}}
\makeatother
\providecommand*\mcitethebibliography{\thebibliography}
\csname @ifundefined\endcsname{endmcitethebibliography}  {\let\endmcitethebibliography\endthebibliography}{}
\begin{mcitethebibliography}{40}
\providecommand*\natexlab[1]{#1}
\providecommand*\mciteSetBstSublistMode[1]{}
\providecommand*\mciteSetBstMaxWidthForm[2]{}
\providecommand*\mciteBstWouldAddEndPuncttrue
  {\def\EndOfBibitem{\unskip.}}
\providecommand*\mciteBstWouldAddEndPunctfalse
  {\let\EndOfBibitem\relax}
\providecommand*\mciteSetBstMidEndSepPunct[3]{}
\providecommand*\mciteSetBstSublistLabelBeginEnd[3]{}
\providecommand*\EndOfBibitem{}
\mciteSetBstSublistMode{f}
\mciteSetBstMaxWidthForm{subitem}{(\alph{mcitesubitemcount})}
\mciteSetBstSublistLabelBeginEnd
  {\mcitemaxwidthsubitemform\space}
  {\relax}
  {\relax}

\bibitem[Wolf \latin{et~al.}(2001)Wolf, Awschalom, Buhrman, Daughton, von Moln{\'a}r, Roukes, Chtchelkanova, and Treger]{wolf2001spintronics}
Wolf,~S.; Awschalom,~D.; Buhrman,~R.; Daughton,~J.; von Moln{\'a}r,~v.~S.; Roukes,~M.; Chtchelkanova,~A.~Y.; Treger,~D. Spintronics: a spin-based electronics vision for the future. \emph{science} \textbf{2001}, \emph{294}, 1488--1495\relax
\mciteBstWouldAddEndPuncttrue
\mciteSetBstMidEndSepPunct{\mcitedefaultmidpunct}
{\mcitedefaultendpunct}{\mcitedefaultseppunct}\relax
\EndOfBibitem
\bibitem[Naber \latin{et~al.}(2007)Naber, Faez, and van~der Wiel]{naber2007organic}
Naber,~W.; Faez,~S.; van~der Wiel,~W.~G. Organic spintronics. \emph{Journal of Physics D: Applied Physics} \textbf{2007}, \emph{40}, R205\relax
\mciteBstWouldAddEndPuncttrue
\mciteSetBstMidEndSepPunct{\mcitedefaultmidpunct}
{\mcitedefaultendpunct}{\mcitedefaultseppunct}\relax
\EndOfBibitem
\bibitem[Sanvito(2011)]{sanvito2011molecular}
Sanvito,~S. Molecular spintronics. \emph{Chemical Society Reviews} \textbf{2011}, \emph{40}, 3336--3355\relax
\mciteBstWouldAddEndPuncttrue
\mciteSetBstMidEndSepPunct{\mcitedefaultmidpunct}
{\mcitedefaultendpunct}{\mcitedefaultseppunct}\relax
\EndOfBibitem
\bibitem[Pandey \latin{et~al.}(2023)Pandey, Sharangi, Sahoo, Mahanta, Mallik, and Bedanta]{pandey2023perspective}
Pandey,~E.; Sharangi,~P.; Sahoo,~A.; Mahanta,~S.~P.; Mallik,~S.; Bedanta,~S. A Perspective on multifunctional ferromagnet/organic molecule spinterface. \emph{Applied Physics Letters} \textbf{2023}, \emph{123}\relax
\mciteBstWouldAddEndPuncttrue
\mciteSetBstMidEndSepPunct{\mcitedefaultmidpunct}
{\mcitedefaultendpunct}{\mcitedefaultseppunct}\relax
\EndOfBibitem
\bibitem[Barraud \latin{et~al.}(2010)Barraud, Seneor, Mattana, Fusil, Bouzehouane, Deranlot, Graziosi, Hueso, Bergenti, Dediu, \latin{et~al.} others]{barraud2010unravelling}
Barraud,~C.; Seneor,~P.; Mattana,~R.; Fusil,~S.; Bouzehouane,~K.; Deranlot,~C.; Graziosi,~P.; Hueso,~L.; Bergenti,~I.; Dediu,~V.; others Unravelling the role of the interface for spin injection into organic semiconductors. \emph{Nature Physics} \textbf{2010}, \emph{6}, 615--620\relax
\mciteBstWouldAddEndPuncttrue
\mciteSetBstMidEndSepPunct{\mcitedefaultmidpunct}
{\mcitedefaultendpunct}{\mcitedefaultseppunct}\relax
\EndOfBibitem
\bibitem[Zhang \latin{et~al.}(2020)Zhang, Guo, Zhu, and Sun]{zhang2020application}
Zhang,~Y.; Guo,~L.; Zhu,~X.; Sun,~X. The application of organic semiconductor materials in spintronics. \emph{Frontiers in Chemistry} \textbf{2020}, \emph{8}, 589207\relax
\mciteBstWouldAddEndPuncttrue
\mciteSetBstMidEndSepPunct{\mcitedefaultmidpunct}
{\mcitedefaultendpunct}{\mcitedefaultseppunct}\relax
\EndOfBibitem
\bibitem[Guo \latin{et~al.}(2019)Guo, Qin, Gu, Zhu, Zhou, and Sun]{guo2019spin}
Guo,~L.; Qin,~Y.; Gu,~X.; Zhu,~X.; Zhou,~Q.; Sun,~X. Spin transport in organic molecules. \emph{Frontiers in Chemistry} \textbf{2019}, \emph{7}, 428\relax
\mciteBstWouldAddEndPuncttrue
\mciteSetBstMidEndSepPunct{\mcitedefaultmidpunct}
{\mcitedefaultendpunct}{\mcitedefaultseppunct}\relax
\EndOfBibitem
\bibitem[Mallik \latin{et~al.}(2018)Mallik, Mattauch, Dalai, Br{\"u}ckel, and Bedanta]{mallik2018effect}
Mallik,~S.; Mattauch,~S.; Dalai,~M.~K.; Br{\"u}ckel,~T.; Bedanta,~S. Effect of magnetic fullerene on magnetization reversal created at the Fe/C60 interface. \emph{Scientific reports} \textbf{2018}, \emph{8}, 5515\relax
\mciteBstWouldAddEndPuncttrue
\mciteSetBstMidEndSepPunct{\mcitedefaultmidpunct}
{\mcitedefaultendpunct}{\mcitedefaultseppunct}\relax
\EndOfBibitem
\bibitem[Mallik \latin{et~al.}(2019)Mallik, Sharangi, Sahoo, Mattauch, Br{\"u}ckel, and Bedanta]{mallik2019enhanced}
Mallik,~S.; Sharangi,~P.; Sahoo,~B.; Mattauch,~S.; Br{\"u}ckel,~T.; Bedanta,~S. Enhanced anisotropy and study of magnetization reversal in Co/C60 bilayer thin film. \emph{Applied physics letters} \textbf{2019}, \emph{115}\relax
\mciteBstWouldAddEndPuncttrue
\mciteSetBstMidEndSepPunct{\mcitedefaultmidpunct}
{\mcitedefaultendpunct}{\mcitedefaultseppunct}\relax
\EndOfBibitem
\bibitem[Sharangi \latin{et~al.}(2021)Sharangi, Pandey, Mohanty, Nayak, and Bedanta]{sharangi2021spinterface}
Sharangi,~P.; Pandey,~E.; Mohanty,~S.; Nayak,~S.; Bedanta,~S. Spinterface-Induced Modification in Magnetic Properties in Co40Fe40B20/Fullerene Bilayers. \emph{The Journal of Physical Chemistry C} \textbf{2021}, \emph{125}, 25350--25355\relax
\mciteBstWouldAddEndPuncttrue
\mciteSetBstMidEndSepPunct{\mcitedefaultmidpunct}
{\mcitedefaultendpunct}{\mcitedefaultseppunct}\relax
\EndOfBibitem
\bibitem[Bairagi \latin{et~al.}(2015)Bairagi, Bellec, Repain, Chacon, Girard, Garreau, Lagoute, Rousset, Breitwieser, Hu, \latin{et~al.} others]{bairagi2015tuning}
Bairagi,~K.; Bellec,~A.; Repain,~V.; Chacon,~C.; Girard,~Y.; Garreau,~Y.; Lagoute,~J.; Rousset,~S.; Breitwieser,~R.; Hu,~Y.-C.; others Tuning the magnetic anisotropy at a molecule-metal interface. \emph{Physical review letters} \textbf{2015}, \emph{114}, 247203\relax
\mciteBstWouldAddEndPuncttrue
\mciteSetBstMidEndSepPunct{\mcitedefaultmidpunct}
{\mcitedefaultendpunct}{\mcitedefaultseppunct}\relax
\EndOfBibitem
\bibitem[Delprat \latin{et~al.}(2018)Delprat, Galbiati, Tatay, Quinard, Barraud, Petroff, Seneor, and Mattana]{delprat2018molecular}
Delprat,~S.; Galbiati,~M.; Tatay,~S.; Quinard,~B.; Barraud,~C.; Petroff,~F.; Seneor,~P.; Mattana,~R. Molecular spintronics: the role of spin-dependent hybridization. \emph{Journal of Physics D: Applied Physics} \textbf{2018}, \emph{51}, 473001\relax
\mciteBstWouldAddEndPuncttrue
\mciteSetBstMidEndSepPunct{\mcitedefaultmidpunct}
{\mcitedefaultendpunct}{\mcitedefaultseppunct}\relax
\EndOfBibitem
\bibitem[Sanvito(2010)]{sanvito2010rise}
Sanvito,~S. The rise of spinterface science. \emph{Nature Physics} \textbf{2010}, \emph{6}, 562--564\relax
\mciteBstWouldAddEndPuncttrue
\mciteSetBstMidEndSepPunct{\mcitedefaultmidpunct}
{\mcitedefaultendpunct}{\mcitedefaultseppunct}\relax
\EndOfBibitem
\bibitem[Sharangi \latin{et~al.}(2022)Sharangi, Mukhopadhyaya, Mallik, Pandey, Ojha, Ali, and Bedanta]{sharangi2022effect}
Sharangi,~P.; Mukhopadhyaya,~A.; Mallik,~S.; Pandey,~E.; Ojha,~B.; Ali,~M.~E.; Bedanta,~S. Effect of fullerene on the anisotropy, domain size and relaxation of a perpendicularly magnetized Pt/Co/C 60/Pt system. \emph{Journal of Materials Chemistry C} \textbf{2022}, \emph{10}, 17236--17244\relax
\mciteBstWouldAddEndPuncttrue
\mciteSetBstMidEndSepPunct{\mcitedefaultmidpunct}
{\mcitedefaultendpunct}{\mcitedefaultseppunct}\relax
\EndOfBibitem
\bibitem[Sun \latin{et~al.}(2010)Sun, Yin, Sun, Guo, Gai, Zhang, Ward, Cheng, and Shen]{sun2010giant}
Sun,~D.; Yin,~L.; Sun,~C.; Guo,~H.; Gai,~Z.; Zhang,~X.-G.; Ward,~T.~Z.; Cheng,~Z.; Shen,~J. Giant magnetoresistance in organic spin valves. \emph{Physical review letters} \textbf{2010}, \emph{104}, 236602\relax
\mciteBstWouldAddEndPuncttrue
\mciteSetBstMidEndSepPunct{\mcitedefaultmidpunct}
{\mcitedefaultendpunct}{\mcitedefaultseppunct}\relax
\EndOfBibitem
\bibitem[Zhang \latin{et~al.}(2014)Zhang, Mizukami, Ma, Kubota, Oogane, Naganuma, Ando, and Miyazaki]{zhang2014spin}
Zhang,~X.; Mizukami,~S.; Ma,~Q.; Kubota,~T.; Oogane,~M.; Naganuma,~H.; Ando,~Y.; Miyazaki,~T. Spin-dependent transport behavior in C60 and Alq3 based spin valves with a magnetite electrode. \emph{Journal of Applied Physics} \textbf{2014}, \emph{115}\relax
\mciteBstWouldAddEndPuncttrue
\mciteSetBstMidEndSepPunct{\mcitedefaultmidpunct}
{\mcitedefaultendpunct}{\mcitedefaultseppunct}\relax
\EndOfBibitem
\bibitem[C{\"o}lle and Br{\"u}tting(2004)C{\"o}lle, and Br{\"u}tting]{colle2004thermal}
C{\"o}lle,~M.; Br{\"u}tting,~W. Thermal, structural and photophysical properties of the organic semiconductor Alq3. \emph{physica status solidi (a)} \textbf{2004}, \emph{201}, 1095--1115\relax
\mciteBstWouldAddEndPuncttrue
\mciteSetBstMidEndSepPunct{\mcitedefaultmidpunct}
{\mcitedefaultendpunct}{\mcitedefaultseppunct}\relax
\EndOfBibitem
\bibitem[Droghetti \latin{et~al.}(2016)Droghetti, Thielen, Rungger, Haag, Gro{\ss}mann, St{\"o}ckl, Stadtm{\"u}ller, Aeschlimann, Sanvito, and Cinchetti]{droghetti2016dynamic}
Droghetti,~A.; Thielen,~P.; Rungger,~I.; Haag,~N.; Gro{\ss}mann,~N.; St{\"o}ckl,~J.; Stadtm{\"u}ller,~B.; Aeschlimann,~M.; Sanvito,~S.; Cinchetti,~M. Dynamic spin filtering at the Co/Alq3 interface mediated by weakly coupled second layer molecules. \emph{Nature Communications} \textbf{2016}, \emph{7}, 12668\relax
\mciteBstWouldAddEndPuncttrue
\mciteSetBstMidEndSepPunct{\mcitedefaultmidpunct}
{\mcitedefaultendpunct}{\mcitedefaultseppunct}\relax
\EndOfBibitem
\bibitem[Dediu \latin{et~al.}(2008)Dediu, Hueso, Bergenti, Riminucci, Borgatti, Graziosi, Newby, Casoli, De~Jong, Taliani, \latin{et~al.} others]{dediu2008room}
Dediu,~V.; Hueso,~L.; Bergenti,~I.; Riminucci,~A.; Borgatti,~F.; Graziosi,~P.; Newby,~C.; Casoli,~F.; De~Jong,~M.~P.; Taliani,~C.; others Room-temperature spintronic effects in Alq 3-based hybrid devices. \emph{Physical Review B} \textbf{2008}, \emph{78}, 115203\relax
\mciteBstWouldAddEndPuncttrue
\mciteSetBstMidEndSepPunct{\mcitedefaultmidpunct}
{\mcitedefaultendpunct}{\mcitedefaultseppunct}\relax
\EndOfBibitem
\bibitem[Xiong \latin{et~al.}(2004)Xiong, Wu, Valy~Vardeny, and Shi]{xiong2004giant}
Xiong,~Z.; Wu,~D.; Valy~Vardeny,~Z.; Shi,~J. Giant magnetoresistance in organic spin-valves. \emph{Nature} \textbf{2004}, \emph{427}, 821--824\relax
\mciteBstWouldAddEndPuncttrue
\mciteSetBstMidEndSepPunct{\mcitedefaultmidpunct}
{\mcitedefaultendpunct}{\mcitedefaultseppunct}\relax
\EndOfBibitem
\bibitem[Tang and VanSlyke(1987)Tang, and VanSlyke]{tang1987organic}
Tang,~C.~W.; VanSlyke,~S.~A. Organic electroluminescent diodes. \emph{Applied physics letters} \textbf{1987}, \emph{51}, 913--915\relax
\mciteBstWouldAddEndPuncttrue
\mciteSetBstMidEndSepPunct{\mcitedefaultmidpunct}
{\mcitedefaultendpunct}{\mcitedefaultseppunct}\relax
\EndOfBibitem
\bibitem[Fukushima and Kaji(2012)Fukushima, and Kaji]{fukushima2012green}
Fukushima,~T.; Kaji,~H. Green-and blue-emitting tris (8-hydroxyquinoline) aluminum (III)(Alq3) crystalline polymorphs: Preparation and application to organic light-emitting diodes. \emph{Organic Electronics} \textbf{2012}, \emph{13}, 2985--2990\relax
\mciteBstWouldAddEndPuncttrue
\mciteSetBstMidEndSepPunct{\mcitedefaultmidpunct}
{\mcitedefaultendpunct}{\mcitedefaultseppunct}\relax
\EndOfBibitem
\bibitem[Jang \latin{et~al.}(2012)Jang, Pernstich, Gundlach, Jurchescu, Richter, \latin{et~al.} others]{jang2012observation}
Jang,~H.-J.; Pernstich,~K.~P.; Gundlach,~D.~J.; Jurchescu,~O.~D.; Richter,~C.; others Observation of spin-polarized electron transport in Alq3 by using a low work function metal. \emph{Applied Physics Letters} \textbf{2012}, \emph{101}\relax
\mciteBstWouldAddEndPuncttrue
\mciteSetBstMidEndSepPunct{\mcitedefaultmidpunct}
{\mcitedefaultendpunct}{\mcitedefaultseppunct}\relax
\EndOfBibitem
\bibitem[Zhan \latin{et~al.}(2010)Zhan, Holmstr{\"o}m, Liz{\'a}rraga, Eriksson, Liu, Li, Carlegrim, Stafstr{\"o}m, and Fahlman]{zhan2010efficient}
Zhan,~Y.; Holmstr{\"o}m,~E.; Liz{\'a}rraga,~R.; Eriksson,~O.; Liu,~X.; Li,~F.; Carlegrim,~E.; Stafstr{\"o}m,~S.; Fahlman,~M. Efficient spin injection through exchange coupling at organic semiconductor/ferromagnet heterojunctions. \emph{Advanced Materials} \textbf{2010}, \emph{22}, 1626--1630\relax
\mciteBstWouldAddEndPuncttrue
\mciteSetBstMidEndSepPunct{\mcitedefaultmidpunct}
{\mcitedefaultendpunct}{\mcitedefaultseppunct}\relax
\EndOfBibitem
\bibitem[Kawahara \latin{et~al.}(2012)Kawahara, Lagoute, Repain, Chacon, Girard, Rousset, Smogunov, and Barreteau]{kawahara2012large}
Kawahara,~S.; Lagoute,~J.; Repain,~V.; Chacon,~C.; Girard,~Y.; Rousset,~S.; Smogunov,~A.; Barreteau,~C. Large magnetoresistance through a single molecule due to a spin-split hybridized orbital. \emph{Nano letters} \textbf{2012}, \emph{12}, 4558--4563\relax
\mciteBstWouldAddEndPuncttrue
\mciteSetBstMidEndSepPunct{\mcitedefaultmidpunct}
{\mcitedefaultendpunct}{\mcitedefaultseppunct}\relax
\EndOfBibitem
\bibitem[Glavic and Bj{\"o}rck(2022)Glavic, and Bj{\"o}rck]{glavic2022genx}
Glavic,~A.; Bj{\"o}rck,~M. GenX 3: the latest generation of an established tool. \emph{Journal of applied crystallography} \textbf{2022}, \emph{55}\relax
\mciteBstWouldAddEndPuncttrue
\mciteSetBstMidEndSepPunct{\mcitedefaultmidpunct}
{\mcitedefaultendpunct}{\mcitedefaultseppunct}\relax
\EndOfBibitem
\bibitem[Galbiati(2015)]{galbiati2015molecular}
Galbiati,~M. \emph{Molecular spintronics: from organic semiconductors to self-assembled monolayers}; Springer, 2015\relax
\mciteBstWouldAddEndPuncttrue
\mciteSetBstMidEndSepPunct{\mcitedefaultmidpunct}
{\mcitedefaultendpunct}{\mcitedefaultseppunct}\relax
\EndOfBibitem
\bibitem[Sakurai \latin{et~al.}(2004)Sakurai, Hosoi, Ishii, Ouchi, Salvan, Kobitski, Kampen, Zahn, and Seki]{sakurai2004study}
Sakurai,~Y.; Hosoi,~Y.; Ishii,~H.; Ouchi,~Y.; Salvan,~G.; Kobitski,~A.; Kampen,~T.; Zahn,~D.; Seki,~K. Study of the interaction of tris-(8-hydroxyquinoline) aluminum (Alq 3) with potassium using vibrational spectroscopy: Examination of possible isomerization upon K doping. \emph{Journal of applied physics} \textbf{2004}, \emph{96}, 5534--5542\relax
\mciteBstWouldAddEndPuncttrue
\mciteSetBstMidEndSepPunct{\mcitedefaultmidpunct}
{\mcitedefaultendpunct}{\mcitedefaultseppunct}\relax
\EndOfBibitem
\bibitem[Davis and Pemberton(2009)Davis, and Pemberton]{davis2009surface}
Davis,~R.~J.; Pemberton,~J.~E. Surface Raman spectroscopy of the interface of tris-(8-hydroxyquinoline) aluminum with Mg. \emph{Journal of the American Chemical Society} \textbf{2009}, \emph{131}, 10009--10014\relax
\mciteBstWouldAddEndPuncttrue
\mciteSetBstMidEndSepPunct{\mcitedefaultmidpunct}
{\mcitedefaultendpunct}{\mcitedefaultseppunct}\relax
\EndOfBibitem
\bibitem[Hubert \latin{et~al.}(1998)Hubert, Sch{\"a}fer, Hubert, and Sch{\"a}fer]{hubert1998domain}
Hubert,~A.; Sch{\"a}fer,~R.; Hubert,~A.; Sch{\"a}fer,~R. Domain Observation and Interpretation. \emph{Magnetic Domains: The Analysis of Magnetic Microstructures} \textbf{1998}, 373--492\relax
\mciteBstWouldAddEndPuncttrue
\mciteSetBstMidEndSepPunct{\mcitedefaultmidpunct}
{\mcitedefaultendpunct}{\mcitedefaultseppunct}\relax
\EndOfBibitem
\bibitem[Hubert and Sch{\"a}fer(1998)Hubert, and Sch{\"a}fer]{hubert1998magnetic}
Hubert,~A.; Sch{\"a}fer,~R. \emph{Magnetic domains: the analysis of magnetic microstructures}; Springer Science \& Business Media, 1998\relax
\mciteBstWouldAddEndPuncttrue
\mciteSetBstMidEndSepPunct{\mcitedefaultmidpunct}
{\mcitedefaultendpunct}{\mcitedefaultseppunct}\relax
\EndOfBibitem
\bibitem[Smith \latin{et~al.}(1960)Smith, Cohen, and Weiss]{smith1960oblique}
Smith,~D.; Cohen,~M.; Weiss,~G.~P. Oblique-incidence anisotropy in evaporated Permalloy films. \emph{Journal of Applied Physics} \textbf{1960}, \emph{31}, 1755--1762\relax
\mciteBstWouldAddEndPuncttrue
\mciteSetBstMidEndSepPunct{\mcitedefaultmidpunct}
{\mcitedefaultendpunct}{\mcitedefaultseppunct}\relax
\EndOfBibitem
\bibitem[Mallik \latin{et~al.}(2019)Mallik, Mohd, Koutsioubas, Mattauch, Satpati, Br{\"u}ckel, and Bedanta]{mallik2019tuning}
Mallik,~S.; Mohd,~A.~S.; Koutsioubas,~A.; Mattauch,~S.; Satpati,~B.; Br{\"u}ckel,~T.; Bedanta,~S. Tuning spinterface properties in iron/fullerene thin films. \emph{Nanotechnology} \textbf{2019}, \emph{30}, 435705\relax
\mciteBstWouldAddEndPuncttrue
\mciteSetBstMidEndSepPunct{\mcitedefaultmidpunct}
{\mcitedefaultendpunct}{\mcitedefaultseppunct}\relax
\EndOfBibitem
\bibitem[Woltersdorf(2004)]{woltersdorf2004spin}
Woltersdorf,~G. Spin-pumping and two-magnon scattering in magnetic multilayers. Ph.D.\ thesis, Sinon Fraser University, 2004\relax
\mciteBstWouldAddEndPuncttrue
\mciteSetBstMidEndSepPunct{\mcitedefaultmidpunct}
{\mcitedefaultendpunct}{\mcitedefaultseppunct}\relax
\EndOfBibitem
\bibitem[Kittel(1948)]{kittel1948theory}
Kittel,~C. On the theory of ferromagnetic resonance absorption. \emph{Physical review} \textbf{1948}, \emph{73}, 155\relax
\mciteBstWouldAddEndPuncttrue
\mciteSetBstMidEndSepPunct{\mcitedefaultmidpunct}
{\mcitedefaultendpunct}{\mcitedefaultseppunct}\relax
\EndOfBibitem
\bibitem[Kalarickal \latin{et~al.}(2006)Kalarickal, Krivosik, Wu, Patton, Schneider, Kabos, Silva, and Nibarger]{kalarickal2006ferromagnetic}
Kalarickal,~S.~S.; Krivosik,~P.; Wu,~M.; Patton,~C.~E.; Schneider,~M.~L.; Kabos,~P.; Silva,~T.~J.; Nibarger,~J.~P. Ferromagnetic resonance linewidth in metallic thin films: Comparison of measurement methods. \emph{Journal of Applied Physics} \textbf{2006}, \emph{99}\relax
\mciteBstWouldAddEndPuncttrue
\mciteSetBstMidEndSepPunct{\mcitedefaultmidpunct}
{\mcitedefaultendpunct}{\mcitedefaultseppunct}\relax
\EndOfBibitem
\bibitem[Heinrich \latin{et~al.}(1985)Heinrich, Cochran, and Hasegawa]{heinrich1985fmr}
Heinrich,~B.; Cochran,~J.; Hasegawa,~R. FMR linebroadening in metals due to two-magnon scattering. \emph{Journal of Applied Physics} \textbf{1985}, \emph{57}, 3690--3692\relax
\mciteBstWouldAddEndPuncttrue
\mciteSetBstMidEndSepPunct{\mcitedefaultmidpunct}
{\mcitedefaultendpunct}{\mcitedefaultseppunct}\relax
\EndOfBibitem
\bibitem[Swindells and Atkinson(2022)Swindells, and Atkinson]{swindells2022interface}
Swindells,~C.; Atkinson,~D. Interface enhanced precessional damping in spintronic multilayers: A perspective. \emph{Journal of Applied Physics} \textbf{2022}, \emph{131}\relax
\mciteBstWouldAddEndPuncttrue
\mciteSetBstMidEndSepPunct{\mcitedefaultmidpunct}
{\mcitedefaultendpunct}{\mcitedefaultseppunct}\relax
\EndOfBibitem
\bibitem[Pan \latin{et~al.}(2017)Pan, Seki, Takanashi, and Barman]{pan2017role}
Pan,~S.; Seki,~T.; Takanashi,~K.; Barman,~A. Role of the Cr buffer layer in the thickness-dependent ultrafast magnetization dynamics of Co 2 Fe 0.4 Mn 0.6 Si Heusler alloy thin films. \emph{Physical Review Applied} \textbf{2017}, \emph{7}, 064012\relax
\mciteBstWouldAddEndPuncttrue
\mciteSetBstMidEndSepPunct{\mcitedefaultmidpunct}
{\mcitedefaultendpunct}{\mcitedefaultseppunct}\relax
\EndOfBibitem
\end{mcitethebibliography}
\end{document}